\documentclass[aip,apl,reprint]{revtex4-1}
\usepackage{graphicx}
\usepackage{dcolumn}
\usepackage{bm}

\usepackage{color}

\begin{document}
\title{dc-switchable and single-nanocrystal-addressable coherent population transfer}

\author{Deniz \surname{Gunceler}}
\email{deniz.gunceler@gmail.com}
\author{Ceyhun \surname{Bulutay}}
\email{bulutay@bilkent.edu.tr}
\affiliation{Department of Physics, Bilkent University, 06800 Bilkent, Ankara, Turkey.}

\begin{abstract}
Achieving coherent population transfer in the solid-state is challenging compared to atomic systems
due to closely spaced electronic states and fast decoherence. 
Here, within an atomistic pseudopotential theory, we offer recipes for the stimulated Raman 
adiabatic passage in embedded silicon and germanium nanocrystals.
The transfer efficiency spectra displays characteristic Fano resonances. 
By exploiting the Stark effect, we predict that transfer can be switched off with a dc voltage. 
As the population transfer is highly sensitive to structural variations, 
with a choice of a sufficiently small two-photon detuning bandwidth, it can be harnessed for addressing 
individual nanocrystals within an ensemble.
\end{abstract}

\pacs{78.67.Bf, 78.67.Hc, 42.50.Hz}

\maketitle 

The control of the dynamics of quantum systems using coherent optical beams lies at the heart of quantum
information technologies.\cite{kral07,saffman10} Among several
alternatives, the stimulated Raman adiabatic passage (STIRAP) offers a certain degree of robustness 
in atomic systems with respect to laser parameter fluctuations.\cite{bergmann98} 
Its solid-state implementation was recently achieved in Pr$^{+3}$:Y$_2$SiO$_5$
crystal,\cite{klein07,wang08} and Tm$^{+3}$:YAG crystal,\cite{alexander08} all at cryogenic temperatures.
Next obvious milestone is to demonstrate STIRAP in nanocrystals (NCs) embedded 
in a host lattice. Compared to rare-earth doped ions in inorganic solids,\cite{klein07,wang08,alexander08} 
NCs bring further flexibility in the design of the functional units by tailoring the physical 
parameters such as material composition, size, shape, and strain, together with the external fields.
However, the challenge with NCs is that the charge degrees of freedom is more susceptible 
to decoherence compared to the atomic or ``impurity'' systems.\cite{takagahara99,forstner03}

In this work we consider silicon and germanium NCs embedded in silica.
Our aim here is to explore from a theoretical standpoint the feasibility as well as
the intricacies of STIRAP in this system. 
To set the stage, first we need to address the constraints imposed by decoherence on our system. 
The ultimate decoherence mechanism is the radiative recombination. 
The typical radiative lifetimes of Si and Ge NCs are in the microsecond range whereas for direct 
band gap semiconductors this is in the nanosecond range.\cite{delerue04} 
Another recombination channel, in case multiple electrons get excited by a strong laser 
pulse, is the Auger process. According to our recent theoretical estimation for the excited-electron
configuration of Auger recombination in Si and Ge NCs, this lifetime 
is in the range of sub-nanoseconds.\cite{sevik08} An even more critical decoherence channel in NCs 
is the acoustic phonon scattering.\cite{takagahara99,forstner03}
For the case of InGaAs quantum dots, Borri \emph{et al.} have demonstrated close to radiative limit 
linewidth at 7~K, corresponding to a dephasing time of 630~ps.\cite{borri01} 
In Si NCs, Sychungov \emph{et al.} have shown that the linewidth can also be as sharp as 
direct band gap materials, reaching 2~meV at 35~K.\cite{sychungov05}
A similar system is the excited Rydberg states\cite{saffman10} of 
phosphorus-doped silicon having a spatial extend of $\sim$10~nm for which 
a dephasing time of $\sim$320~ps is very recently predicted.\cite{greenland10}
Guided by these reports, we aim for a complete STIRAP in less than 300~ps so that 
at sufficiently low temperatures of a few Kelvins this can beat the decoherence clock in Si and Ge NCs.
Admittedly, this is a cautiously optimistic estimate, nevertheless a worse case can still be accommodated 
by further scaling the pulse widths and laser powers accordingly; thanks to high-field 
tolerance of silica embedded NCs.

Our theoretical model involves the atomistic description of the system 
within a supercell geometry of several thousand atoms most of which are the 
surrounding matrix atoms. Initially, nearly spherical NCs in $C_{3v}$ point
symmetry are considered, and in the final part consequences of shape deformation 
are discussed. The local potential is represented as a superposition of 
screened semiempirical pseudopotentials of the constituent atoms;\cite{bulutay07} 
the spin-orbit interaction is particularly included, since this coupling among 
closely spaced levels can potentially affect the selection rules and hence the 
transfer efficiency. A dc electric field is also accounted nonperturbatively 
for the Stark field analysis. The excitonic effects are ignored as the confinement 
energy dominates for small NCs.\cite{forstner03}
To solve the single-particle Schr\"odinger equation for such a large number 
of atoms with sufficient accuracy up to the highly excited states, we make use 
of the linear combination of bulk bands approach.\cite{wang99}

The population transfer is built on this atomistic electronic structure 
as shown in the insets of Fig.~1. The electric dipole coupling is used for 
the interaction with the pump and Stokes beams. Unlike atomic systems, in the case of NCs 
we have to consider multiple intermediate\cite{carrol92,vitanov99} and 
final states. On the other hand we assume a single initial state, namely,
the highest occupied molecular orbital (HOMO). 
However, by selecting the interaction parameters  accordingly (see Table~I) 
we have assured that the maximum probability of finding a sub-HOMO electron 
in the conduction band is quite negligible. Our computations show that 
had the interaction individually 
involved any such sub-HOMO electron as the initial state, its transfer 
probability to the conduction band would have an upper bound of $10^{-9}$ \%.
Finally, since the intermediate state is not populated 
for the case of an ideal STIRAP, we neglect many-body effects due to 
Pauli blocking.

\begin{figure}
\includegraphics[width=8.5 cm]{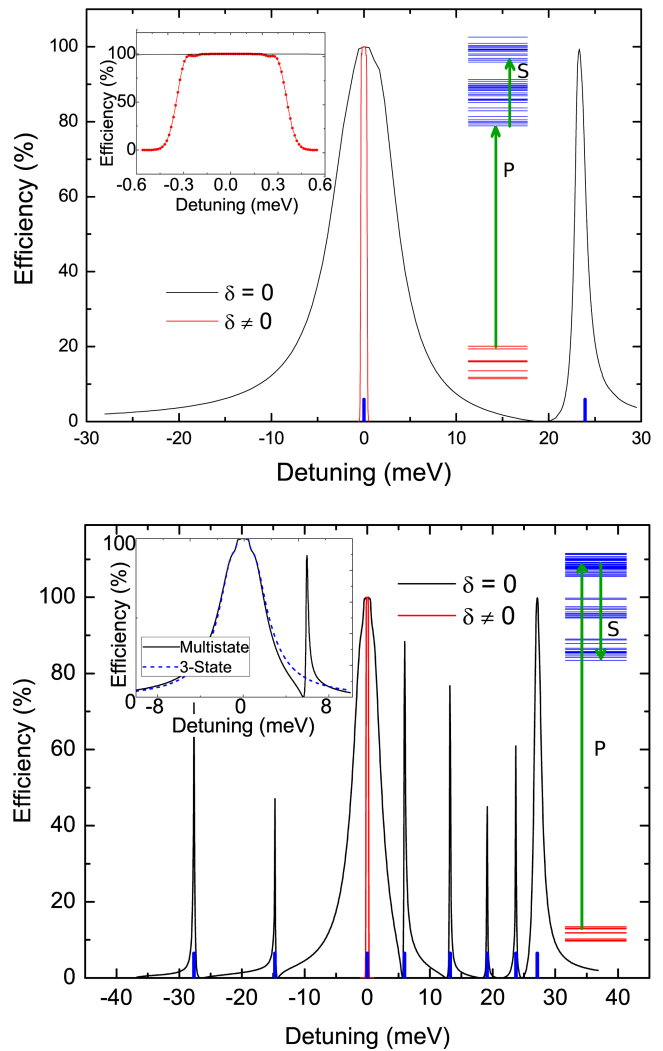}
\caption{\label{fig1} (Color online) The population transfer efficiency as the pump laser 
is detuned for the 1.5~nm Ge NC (top) and the 2.1~nm Si NC (bottom), with ($\delta=0$) 
and without two-photon resonance ($\delta\neq0$). The left inset in the upper graph 
is a close-up for the central peak. The vertical blue lines on the detuning axis show the 
energies of the intermediate states. Insets on the right show the electronic 
states and laser energies.}
\end{figure}

We base our discussions on a spherical 2.1~nm diameter Si NC and a 1.5~nm Ge NC, 
purposely selected because of their close band gaps of 2.74~eV and 2.80~eV, respectively.
Considering the oscillator strength of the transitions, we adopt different 
schemes for the two NCs (see right insets in Fig.~\ref{fig1}). 
For 1.5~nm Ge NC the HOMO to lowest unoccupied molecular orbital (LUMO) 
transition is quite strong, therefore, we utilize the ladder scheme where 
the electron is transferred from the HOMO state to the LUMO+28 state (0.837~eV above LUMO) 
via the LUMO state. For 2.1~nm Si NC, we utilize a $\Lambda$ scheme where the electron
is transferred from the HOMO state to the LUMO state via the LUMO+30 state (0.937~eV above LUMO). 
For the pump and Stokes pulses, we use counterintuitively time-ordered Gaussian profiles\cite{bergmann98} 
giving rise to equal peak Rabi frequencies with values 0.35 and 0.55~THz for Si and Ge NCs, 
respectively. Both pulses are linearly polarized but in certain directions chosen to optimize the transfer. 
We refer to Table~I for the other laser parameters.
 
\begin{table}[h]
\caption{Laser parameters optimized for STIRAP for the 2.1~nm Si, 
and 1.5~nm Ge NCs. The incident electric fields are specified for 
free-space medium. Delay refers to time between the peaks 
of the Stokes and pump pulses both with Gausian profiles.}
\begin{tabular}{c c c c c}
\hline
\hline
 & E-field & Wavelength & FWHM & Delay\\
 & (MV/cm)  &  (nm)     & (ps) & (ps)\\
\hline
Si NC & & & & \\
\begin{tabular}{c} Pump \\ Stokes \end{tabular} & 
\begin{tabular}{c} 1.5569 \\ 0.058954 \end{tabular} & 
\begin{tabular}{c} 336.9 \\ 1323 \end{tabular} & 90 & 75\\
\hline
Ge NC & & & & \\
\begin{tabular}{c} Pump \\ Stokes \end{tabular} & 
\begin{tabular}{c} 0.32325 \\ 0.92810 \end{tabular} & 
\begin{tabular}{c} 442.4 \\ 1481 \end{tabular} & 60 & 50\\
\hline
\hline
\end{tabular}
\end{table}

The transfer efficiency is plotted in Fig.~\ref{fig1}. 
In all the figures, detuning refers to that of the pump pulse; in the case of 
two-photon resonance, the Stokes pulse is assumed to be detuned by 
the same amount from its targeted transition. 
In the other case, the Stokes pulse is kept resonant. The first observation we make is that
the transfer efficiency is superior when the two-photon resonance
condition is satisfied. This fact is well 
documented in literature.\cite{bergmann98} When the system has two-photon detuning, 
the central peak shrinks down into a plateau, where the transfer efficiency is essentially constant,
and suddenly drops to zero (see upper inset). 

Another noteworthy aspect of Fig.~\ref{fig1} is the presence
of side peaks. The detuning values for these peaks coincide with 
the resonance condition restored with a neighboring intermediate state,
marked by the blue vertical bars in Fig.~1. As the laser parameters were 
optimized for the central peak and not for the neighbors,
these peaks are usually not as tall or wide.
Also since there are no states immediately below LUMO 
(ladder scheme), for the 1.5~nm Ge NC there are no such peaks for negative detuning.
The peaks display the asymmetrical well-known Fano 
lineshape.\cite{fano61,miroshnichenko10} In a similar context, 
this was observed in the tunelling-induced 
transparency in quantum well intersubband transitions.\cite{schmidt97}
It arises from two paths interfering with opposite phase on 
one of the two sides of the resonance. This is illustrated  
in the lower left inset of Fig.~\ref{fig1}: For small detunings the 
calculations made without considering any neighbors agree very well 
with the full calculation. However, as soon as 
there is enough detuning to transfer the electron through one of the 
neighboring states, the neighbors-removed treatment cannot reproduce 
the dip right before the second peak in the solid line which occurs due to 
the interference between the chosen intermediate state and its neighbor.

\begin{figure}
\includegraphics[width=8.5 cm]{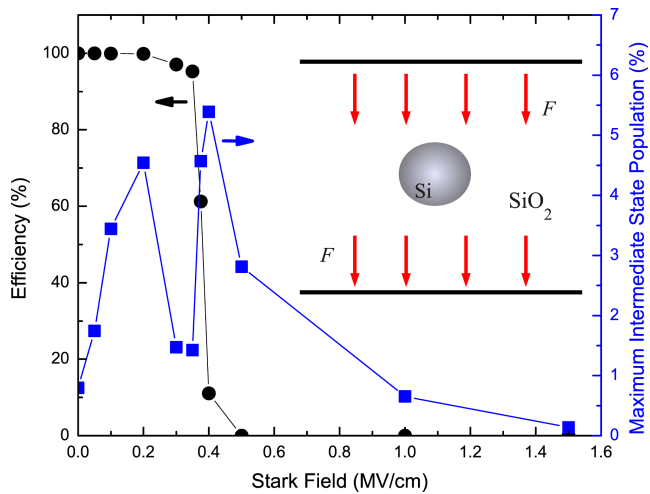}
\caption{\label{fig2} (Color online) Effect of dc Stark field on the overall 
transfer efficiency and maximum intermediate state population for the 2.1~nm 
Si NC. Lines are only to guide the eyes.}
\end{figure}

Next, in Fig.~\ref{fig2}, we investigate the effect of an external dc electric field 
(see the inset). Note that we quote the \emph{matrix} dc electric 
fields, viz., outside the NC in the embedding region of silica. 
Here, for the 2.1~nm Si NC, we observe that for small fields, the transfer 
efficiency is not affected, but the intermediate population pile-up increases.  
After a critical field of 0.35~MV/cm, transfer efficiency rapidly drops to zero. 
To identify its origin, we first assured that the system is robust against changes in the dipole 
matrix elements, hence Rabi frequencies are not significantly altered by the dc field.
On the other hand, NC energy levels undergo significant Stark shifts, the 
valence states being more so compared to conduction states, as 
revealed by our recent work.\cite{bulutay10} We checked that in this field range 
it does not give rise to a level crossing between HOMO and the lower-lying states. 
Hence, the primary mechanism responsible for this switching is the Stark shift-induced two-photon 
detuning, something STIRAP is very sensitive to. For the zero-field case, 
a detuning of 0.3~meV is enough to destroy STIRAP. 
This value of Stark shift is reached at 0.5~MV/cm, after which the population transfer is quenched.

\begin{figure}
\includegraphics[width=8.5 cm]{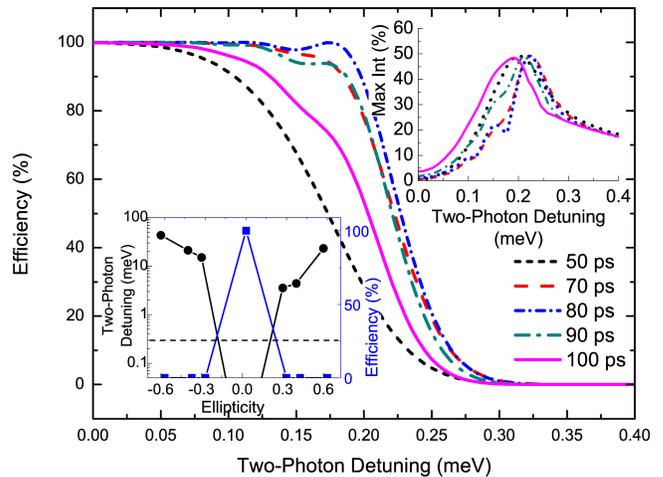}
\caption{\label{fig3} (Color online) Two-photon detuning versus transfer efficiency 
for different time delays for the 2.1~nm Si NC. Upper 
inset: the maximum probability of finding the electron in 
any of the intermediate states throughout the transfer process. Lower inset: 
the effect of NC ellipticity on the two-photon detuning and transfer efficiency 
(Lines are guides to the eyes);
the horizontal dashed line marks the critical 0.3~meV two-photon detuning.}
\end{figure}

The highly critical two-photon bandwidth is controllable by the time delay between Stokes 
and pump pulses which is illustrated in Fig.~\ref{fig3}. As the overlap between these 
pulses is reduced the two-photon bandwidth first increases up to a 
delay of 80~ps, beyond which it retracts back, as expected from the fundamental 
principles of STIRAP that demands a non-zero overlap between the two pulses.\cite{bergmann98}
The upper inset shows the build-up of the intermediate-state population away from the 
two-photon resonance. 

Finally, we focus on the NC's structural sensitivity. 
Starting with its shape, we consider NCs of the same number of atoms 
(hence, same volume under zero strain) 
but with different asphericities as quantified by the ellipticity parameter, $e$. 
Denoting $a$ and $b$ as the equatorial radii and $c$ as the polar radius, 
for oblate spheroids ($a=b>c$) $e=\sqrt{1-c^2/a^2}$, whereas for prolate spheroids 
($a=b<c$) we \emph{define} it to be \emph{negative} as $e=-\sqrt{1-a^2/c^2}$.
The lower inset of Fig.~\ref{fig3} vividly displays the fact that the transfer is lost as the shape 
of the NC is deformed from the originally targeted geometry to which STIRAP 
was optimized (here spherical). The repositioning of only two NC surface atoms ($e$=0.3 case)
is enough to displace the electronic states away from the tolerable two-photon detuning window. 
Likewise, we observed that an incremental change in the size of the NC by including 
the next shell of atoms (not shown) results in a similar loss of transfer. These indicate 
that practically the intended STIRAP will be locked only to the \emph{single} NC that it is tuned to.

In conclusion, we provide a theoretical insight for STIRAP in small Si and Ge NCs. 
Due to dense electronic states it displays a train of Fano resonances. The transfer can be abruptly 
switched off with a dc voltage by introducing Stark shift that sufficiently detunes the 
two-photon resonance. Finally, we demonstrate the sensitivity of the transfer efficiency 
with respect to the structure of the NC which can be instrumental in addressing a single 
NC among an ensemble having inherent size, shape and even local strain fluctuations.

The partial support from the European FP7 Project UNAM-Regpot Grant No. 203953 is acknowledged.

\end{document}